\documentclass[epj,nopacs]{svjour}

\usepackage{graphicx,amsmath,amssymb}

\newcommand{\eg}{e.g., }
\newcommand{\Em}{E_{\rm m}}
\newcommand{\Es}{E_{\rm s}}
\newcommand{\Gammaf}{\Gamma_{\rm f}}

\newcommand{\ie}{i.e., }
\newcommand{\lambdac}{\lambda_{\rm c}}
\newcommand{\omegaf}{\omega_{1{\rm f}}}
\newcommand{\omegam}{\omega_{\rm m}}

\newcommand{\pd}{\partial}
\newcommand{\Pzc}{P_{0{\rm c}}}
\newcommand{\qc}{q_{\rm c}}
\newcommand{\qf}{q_{\rm f}}
\newcommand{\qmax}{q_{\rm max}}
\newcommand{\rhom}{\rho_{\rm m}}
\newcommand{\rhos}{\rho_{\rm s}}
\newcommand{\tK}{\tilde{K}}
\newcommand{\tu}{\tilde{u}}
\newcommand{\tv}{\tilde{v}}
  
\begin{document}

\title{Parametric excitation of wrinkles in elastic sheets on elastic and
  viscoelastic substrates}

\titlerunning{Parametric resonance of supported sheets}

\author{Haim Diamant}

\institute{School of Chemistry, and Center for Physics and Chemistry
  of Living Systems, Tel Aviv University, Tel Aviv 6997801, Israel}

\date{\today}

\abstract{Thin elastic sheets supported on compliant media form
  wrinkles under lateral compression. Since the lateral pressure is
  coupled to the sheet's deformation, varying it periodically in time
  creates a parametric excitation. We study the resulting parametric
  resonance of wrinkling modes in sheets supported on semi-infinite
  elastic or viscoelastic media, at pressures smaller than the
  critical pressure of static wrinkling. We find distinctive behaviors
  as a function of excitation amplitude and frequency, including (a) a
  different dependence of the dynamic wrinkle wavelength on sheet
  thickness compared to the static wavelength; and (b) a discontinuous
  decrease of the dominant wrinkle wavelength upon increasing
  excitation frequency at sufficiently large pressures. In the case of
  a viscoelastic substrate, resonant wrinkling requires crossing a
  threshold of excitation amplitude. The frequencies for observing
  these phenomena in relevant experimental systems are of the order of
  a kilohertz and above. We discuss experimental implications of the
  results.}

\maketitle

\section{Introduction}
\label{sec_intro}

Wrinkling is one of the common deformation patterns which thin elastic
sheets form when subjected to lateral compression
\cite{Cerda2003,Genzer2006,Davidovitch2011}. In many cases wrinkles
appear when the sheet is supported on a softer substrate, a scenario
which is relevant to a range of applications (\eg coatings, paints)
and naturally occurring structures (\eg skin and tissue
linings). Studies have been directed more recently at {\em active}
wrinkling
\cite{Pocivavsek2018,Pocivavsek2019,Nath2020,Lin2020,Wen2020}. The
interplay between the topography of supported thin sheets and their
delamination off the support
\cite{Vella2009,Mei2011,Hohfeld2015,Oshri2018,Oshri2020} suggests
active wrinkling as an anti-fouling strategy adopted by Nature and
mimicked in man-made systems
\cite{Bixler2012,Pocivavsek2019,Nath2020,Wen2020}. These studies of
active wrinkling have considered static or quasi-static wrinkles,
arising from mechanical equilibrium at pressures exceeding the static
flat-to-wrinkle transition. The dynamic effects considered in those
studies \cite{Pocivavsek2019,Wen2020} are due to low-frequency (below
1~Hz) actuations, where the wrinkles follow the external stimulus
quasi-statically.

Works going beyond the quasi-static limit addressed the time evolution
of the flat-to-wrinkle transition in sheets supported on viscous
\cite{Sridhar2001,Huang2002} and viscoelastic \cite{Huang2005}
media. Dynamic wrinkles have been studied in two additional
scenarios. The first is the formation of radial wrinkles in thin
sheets upon impact of a rigid object
\cite{Vermorel2009,Vandenberghe2016,Box2019,Ghanem2019}. In another
scenario a slender body in contact with a liquid is compressed by a
progressively increasing lateral pressure
\cite{Kodio2017,Chopin2017,Box2020}. Unlike static wrinkles, whose
wavelength is determined by a competition between two restoring forces
(\eg bending of the sheet and deformation of the substrate), those
short-time dynamic wrinkles arise from an interplay of a restoring
force and inertia or viscous stresses in the substrate, resulting in a
wavelength that increases with time. Finally, dynamic control of
wrinkle wavelength and pattern was demonstrated in supported sheets
under changing temperature and solvent diffusion
\cite{Vandeparre2010}.

The present work investigates a different phenomenon, where periodic
forcing and inertia take a supported sheet out of plane through a
mechanism of parametric resonance \cite{LLmechanics}. Parametric
resonance suggests itself naturally for compressed sheets, because the
actuating pressure produces a force that depends on the sheet's
out-of-plane deformation.

The investigated dynamics involves a combination of five factors: the
sheet's bending elasticity and inertia, and the substrate's
elasticity, viscosity, and inertia. Section~\ref{sec_scales} is
devoted, therefore, to heuristic consideration of the relevant scales
and dominant mechanisms. In addition, to reduce the complexity of the
analysis, we will employ along the way several simplifying assumptions
while trying not to compromise the qualitative physical significance
of the results.  In sect.~\ref{sec_model} we present the model and the
general equations of motion which are common to the more specific
cases that follow. Section \ref{sec_results} presents results for a
sheet supported on two types of substrate\,---\,an elastic substrate
(sect.~\ref{sec_elastic}) and a viscoelastic one
(sect.~\ref{sec_ve}). Due to the complexity of the problem we give in
the main text the key steps of the derivations and their results. The
detailed calculations are found in the Supplementary Material
\cite{suppl}. In sect.~\ref{sec_discuss} we summarize the predictions
for experiments, compare the resonant wrinkling with other
dynamic-wrinkling scenarios, and describe potential extensions of the
theory.

\section{Relevant scales}
\label{sec_scales}

Let us examine the relevant scales and dominant mechanisms of the
suggested phenomenon.  As mentioned above, five physical mechanisms
at play: (a) the substrate's elasticity, characterized by a
shear modulus $G$; (b) the sheet's rigidity, characterized by a
bending modulus $B$; (c) the substrate's inertia, characterized by a
three-dimensional (3D) mass density $\rhom$; (d) the sheet's inertia,
characterized by a 2D mass density $\rho=\rhos h$, where $\rhos$ is
the sheet's 3D mass density and $h$ its thickness; (e) in the case of
a viscoelastic medium, the substrate's viscosity $\eta$.

Statically, the competition between the rigidities of the sheet and
supporting medium gives rise to an intrinsic length which determines
the wavelength of static wrinkles \cite{Cerda2003,Genzer2006}. For a
semi-infinite elastic substrate, the competition between (a) and (b)
above gives the intrinsic length as $\lambdac\sim(B/G)^{1/3}$
\cite{Groenewold2001}. In terms of the Young moduli of the sheet and
medium, $\Es$ and $\Em$, it can be rewritten as $\lambdac\sim
h(\Es/\Em)^{1/3}$. Thus, sheets that are orders of magnitude stiffer
than the medium are required to obtain wrinkles with wavelength
appreciably larger than $h$.

Dynamically, for a given length scale $q^{-1}$, the balance between
one of the restoring forces and one of the inertial effects determines
(by dimensional analysis) a characteristic frequency. Each of these
balances gives the frequency--wavenumber relation for a limiting
resonance mechanism. Balancing (a) and (c) above gives $\omega_{\rm
  ac} \sim (G/\rhom)^{1/2} q$; this is the relation for Rayleigh waves
on the surface of a sheet-free medium \cite{LLelasticityRayleigh}. The
combination of (b) and (c) gives $\omega_{\rm bc} \sim
(B/\rhom)^{1/2}q^{5/2}$. Taking (a) and (d), we find $\omega_{\rm ad}
\sim (G/\rho)^{1/2}q^{1/2}$. Finally, (b) and (d) give $\omega_{\rm
  bd} \sim (B/\rho)^{1/2}q^2$; this is the relation for bending waves
along a substrate-free sheet.

Assuming $h\ll q^{-1}\sim\lambdac$, one finds $\omega_{\rm ac} \sim
\omega_{\rm bc} \ll \omega_{\rm ad} \sim \omega_{\rm bd}$. This
implies that the dominant inertial effect usually comes from the
substrate rather than the sheet. Hence, although the theory formulated
below accounts for the inertia of both components, we will
subsequently concentrate on the limit of substrate-dominated
inertia. In this limit we expect a crossover in the relation between
actuation frequency and actuated wavenumber, around
$q\sim\lambdac^{-1}$, from $\omega\sim\omega_{\rm ac}\sim q$ to
$\omega\sim \omega_{\rm bc} \sim q^{5/2}$.\footnote{The opposite limit, of
sheet-dominated inertia, is analyzed in the Supplementary Material
\cite{suppl}. In this limit we expect a crossover from $\omega\sim\omega_{\rm
  ad}\sim q^{1/2}$ to $\omega\sim\omega_{\rm bd}\sim q^2$.}

Thus the frequency $\omegam \sim (G/\rhom)^{1/2}/\lambdac$, obtained
from $\omega_{\rm ac}$ or $\omega_{\rm bc}$ for $q\sim\lambdac^{-1}$,
sets the scale for the actuation frequency required to excite wrinkles
of wavelength $\sim\lambdac$. For $G>10^3$~Pa and $\lambdac<1$~mm, we
get $\omegam>10^3$~Hz. Such frequencies probably lie outside the range
of natural scenarios but are experimentally relevant.

In the case of a viscoelastic substrate, for the viscous damping to be
appreciable, we need $\eta > G/\omegam \sim
(G\rhom)^{1/2}\lambdac$. With the bounds above this gives
$\eta>1$~Pa\,s, \ie more than $10^3$ times the viscosity of water.

These conclusions are borne out by the detailed analysis that follows.

\section{Model}
\label{sec_model}

\subsection{The system}

\begin{figure}
  \centerline{ \includegraphics[width=0.5\textwidth]{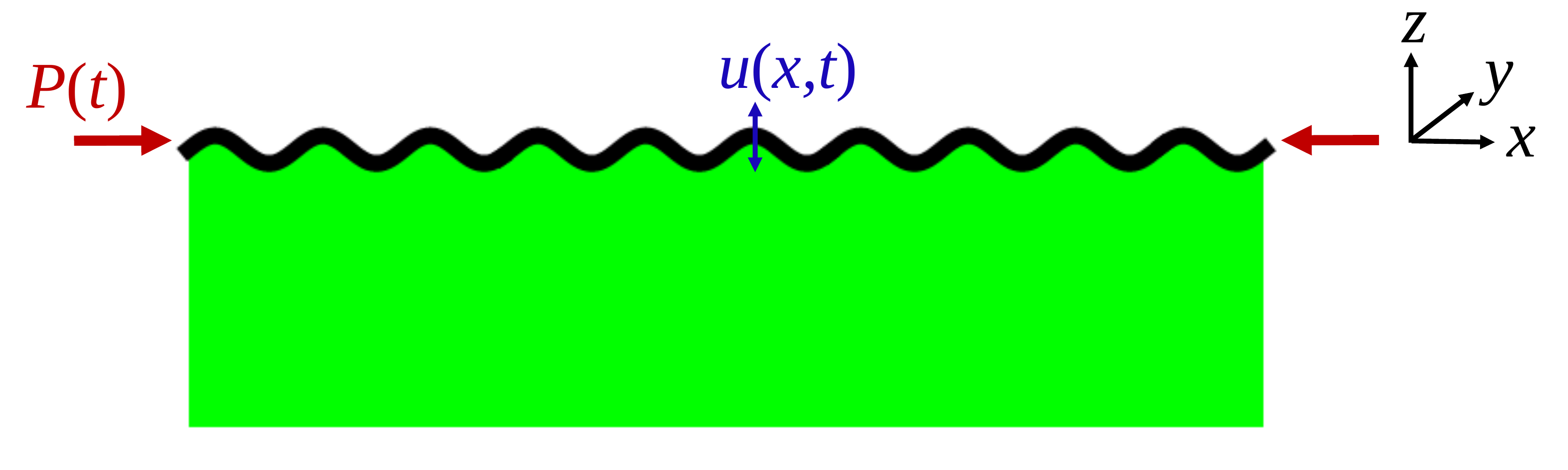}}
  \caption{Schematic view of the system.}
  \label{fig:scheme}
\end{figure}

We consider a thin elastic sheet attached to the surface of a
(visco)elastic medium. The sheet, lying at rest on the $z=0$ plane, is
assumed to be incompressible, infinite, and made of a much stiffer
material than the supporting medium. The medium occupies the region
$z\in(-\infty,0)$. The sheet is compressed unidirectionally, along the
$x$ axis, by a time-dependent actuating pressure (force per unit
length) $P(t)$. It can deform on the $xz$ plane from $z=0$ to
$z=u(x,t)$. See fig.~\ref{fig:scheme}. We assume $|\pd_x u|\ll 1$ and
construct the leading-order (linear) model. Within this approximation
the extension from a one-dimensional surface deformation $u(x,t)$ to a
two-dimensional one, $u(x,y,t)$, is simple, and we restrict the
discussion to 1D for brevity.

\subsection{Equations of motion}
\label{sec_eqs}

Both sheet and medium respond to the surface deformation $u(x,t)$.
The sheet experiences a restoring normal force per unit area due to
bending and the lateral compression,
\begin{equation}
  F_{\rm s}(x,t) = -B u'''' - P(t) u'',
  \label{Fs}
\end{equation}
where a prime denotes an $x$-derivative. We take the actuating
pressure to be
\begin{equation}
  P(t) = P_0 + P_1\cos(\omega_1 t),
  \label{Pt}
\end{equation}
where $P_0$ is the static pressure, $P_1$ the actuation amplitude, and
$\omega_1$ the actuation frequency.

The normal force per unit area which the medium experiences at its
surface is given by the general linear response,
\begin{equation}
  F_{\rm m}(x,t) = \int_{-\infty}^t dt' \int_{-\infty}^\infty dx'
  K(x-x',t-t') u(x',t').
\label{Fm}
\end{equation}
The kernel $K(x,t)$ encodes the effect of the medium's spatial and
temporal response on normal stresses at its surface. In Fourier
space,\\ $\tK(q,\omega) \equiv \int_{-\infty}^\infty dt
\int_{-\infty}^\infty dx e^{iqx - i\omega t} K(x,t)$ is a complex
function arising from the medium's viscoelasticity and inertia. We
will assume a function of the form
\begin{equation}
  \tK(q,\omega) \simeq K_0(q) + i\omega K_1(q) - \omega^2 K_2(q).
  \label{Kgeneral}
\end{equation}
The first and third terms are motivated by the surface response of an
elastic medium in {\em both limits} of low and high frequency (see
below). These time-reversible responses relate to the elastic
restoring force (first term), and the substrate's inertia (third
term). The second, time-irreversible term corresponds to the viscous
component of the response. We assume for simplicity that, within the
relevant frequency range, the viscous coefficient $K_1$ does not
change with frequency (\ie the viscoelasticity is dominated by a
single relaxation process with rate $K_0/K_1$). Note that
eq.~(\ref{Kgeneral}) can be obtained by expanding $\tK$ in a small
range of frequencies around any given frequency.

The equation of motion for the sheet's deformation is
\begin{equation}
  \rho\ddot u = F_{\rm s} - F_{\rm m},
  \label{motion0}
\end{equation}
where $\rho\equiv\rhos h$ is the sheet's mass per unit area, and a dot
denotes a time derivative. Using eqs.~(\ref{Fs})--(\ref{motion0})
while applying a spatial Fourier transform, $\tilde{f}(q,t) \equiv
\int_{-\infty}^\infty dx e^{iqx} f(x,t)$, turns the equation of motion
into
\begin{equation}
  [\rho + K_2(q)] \ddot{\tu} + K_1(q)\dot{\tu} + [Bq^4 - P(t) q^2 + K_0(q)]\tu = 0.
\label{motion1}
\end{equation}
The transformation
\begin{equation}
  \tv \equiv \tu e^{-[K_1/(2(\rho+K_2))] t}
\label{vu}
\end{equation}
eliminates the friction term, yielding
\begin{equation}
  (\rho + K_2)\ddot{\tv} + [Bq^4 - P(t) q^2 + K_0 - K_1^2/(4(\rho+K_2))]\tv = 0.
\label{motion2}
\end{equation}
We rewrite eq.~(\ref{motion2}) as
\begin{equation}
  \ddot{\tv} + \omega_0^2[1 + a\cos((2\omega_0 + \epsilon)t)] \tv = 0,
\label{motionresonance}
\end{equation}
where
\begin{eqnarray}
  \omega_0^2(q) &\equiv& \frac{1}{\rho+K_2}
    \left( Bq^4 - P_0q^2 + K_0 - \frac{K_1^2}
         {4(\rho+K_2)} \right), \nonumber\\
         a(q) &\equiv& - \frac{P_1 q^2}{\omega_0^2},
 \label{resonanceparameters}\\
  \epsilon(q) &\equiv& \omega_1 - 2\omega_0(q). \nonumber
\end{eqnarray}

The problem has been transformed into an analogous chain of
independent, parametrically actuated oscillators, with intrinsic
frequencies $\omega_0(q)$, actuation amplitudes $a(q)$, and detuning
parameters $\epsilon(q)$. We see in eq.~(\ref{resonanceparameters})
that increasing the static pressure $P_0$ weakens the `spring
constant' $\omega_0^2$. For the analogy to work we must have
\begin{equation}
  \omega_0^2(q) > 0,
\end{equation}
and `oscillators' (modes) $q$ which do not satisfy it are
damped. Further, from the known solution to the classical problem of
parametric resonance \cite{LLmechanics}, we infer the condition for
instability (\ie exponentially growing amplitude $\tu(q,t)$), to
leading order in the actuation $a$,
\begin{equation}
  \Gamma^2(q) \equiv \frac{1}{4} a^2\omega_0^2 - \frac{K_1^2}{(\rho+K_2)^2} > 0.
\label{instability0}
\end{equation}
This is the squared rate of amplitude growth. The fastest growing mode
$\qf$ is the one which maximizes $\Gamma(q)$. The allowed detuning for
each `oscillator' $q$, \ie the actuation frequency range providing
resonance, is obtained from the inequality $\epsilon^2(q) <
\Gamma^2(q)$. To simplify the discussion, we will assume perfect
tuning,
\begin{equation}
  \epsilon=0,\ \ \ \  \omega_1=2\omega_0(q).
\label{zero_detuning}
\end{equation}
Thus, by ``unstable band'' we will refer simply to the set of
tuned `oscillators' (\ie range of $q$) for which $\Gamma^2(q)>0$.

\section{Results}
\label{sec_results}

\subsection{Elastic substrate}
\label{sec_elastic}

The kernel $K(x-x',t-t')$ gives the nonlocal time-dependent force
density, acting at a point on the medium's surface at a certain time,
in response to a normal surface displacement occurring elsewhere at a
different time. For a semi-infinite elastic medium its Fourier
transform was calculated by Lamb \cite{Lamb},
\begin{equation}
  \tK(q,\omega) = \frac{4G^2|q|^3}{\rhom\omega^2} \left[
    \left( 1 - \frac{\rhom\omega^2}{Gq^2} \right)^{1/2} -
    \left( 1 - \frac{\rhom\omega^2}{2Gq^2} \right)^2 \right],
\label{KLamb}
\end{equation}
where $G$ is the medium's shear modulus, and we assume for simplicity
an incompressible medium. In both limits of low and high frequency
this expression reduces to the form given by eq.~(\ref{Kgeneral}),
with $K_1=0$, and
\begin{subequations}
  \begin{equation}
    \omega\ll (G/\rhom)^{1/2}|q|:\ \ \ K_0=2G|q|,\ \ K_2=\frac{3\rhom}{2|q|},
  \end{equation}
  \begin{equation}
    \omega\gg (G/\rhom)^{1/2}|q|:\ \ \ K_0=4G|q|,\ K_2=\frac{\rhom}{|q|}.
\label{K_elastic}
  \end{equation}
\label{K_elastic_both}
\end{subequations}
Thus the limits of high and low frequency differ by just numerical
prefactors.\footnote{Note that at intermediate frequencies this kernel
  describes a more complex response, including imaginary (yet still
  time-reversible) terms.} \footnote{In the static limit ($\omega=0$)
  one recovers the result derived from the Boussinesq problem
  \cite{LLelasticity}, $\tK(q,0)=2G|q|$.}

The two regimes defined in eq.~(\ref{K_elastic_both}) can be rewritten
as $\omega/\omegam\ll \lambdac q$ and $\omega/\omegam\gg \lambdac
q$. In the present problem, however, the frequency (of actuation) and
the (excited) wavenumber are inter-related. As we shall see shortly,
for $\omega\ll\omegam$ we get $\lambdac q\sim \omega/\omegam$, and for
$\omega\gg\omegam$, $\lambdac q\ll \omega/\omegam$; namely, the first
limit never strictly holds. Hence, we will assume the second limit and
use eq.~(\ref{K_elastic}). Since the two behaviors are essentially the
same up to constant prefactors, this choice should not have a
qualitative effect. We return to this point in sect.~\ref{sec_discuss}.

To make the expressions concise, we hereafter use $B$ as the unit of
energy, $(B/\hat{G})^{1/3}\sim\lambdac$ as the unit of length, and
$(\rhom/\hat{G})^{1/2}(B/\hat{G})^{1/3} \sim \omegam^{-1}$ as the unit
of time. We choose to multiply $G$ by a numerical prefactor,
$\hat{G}=2G$, such that the static wrinkle wavenumber will turn up
equal to $1$. The rescaling allows us to set
$B=\hat{G}=\rhom=1$. The 2D pressure is then measured in units of
$B^{1/3}\hat{G}^{2/3}$. (In sect.~\ref{sec_experiment} we will rewrite
the most relevant expressions in dimensional form.)

Substituting eq.~(\ref{K_elastic}) in eqs.~(\ref{resonanceparameters})
and (\ref{instability0}) gives $\omega_0$ and $\Gamma$ for the case of
an elastic substrate, including the inertia of both substrate and
sheet. However, based on the estimates in sect.~\ref{sec_scales}, and
to simplify the results, we hereafter neglect the sheet's
inertia. Setting $\rho\rightarrow 0$ in these equations gives
\begin{eqnarray}
  \omega_0^2(q) &=& q^2( q^3 - P_0 q + 2),
\label{omega0_elastic} \\
  \Gamma^2(q) &=& \frac{P_1^2 q^2}{4(q^3 - P_0 q + 2)}.
\label{Gamma_elastic}
\end{eqnarray}
Static wrinkling appears when $\omega_0=0$. This occurs at the critical
pressure and wavenumber
\begin{equation}
  \Pzc = 3,\ \ \ \ \qc = 1.
\end{equation}

For $P_0<\Pzc$ we have $\omega_0^2(q)>0$ and $\Gamma^2(q)>0$ for all
$q$ regardless of $P_1$. Thus all wrinkling modes $q$ are oscillatory
and will resonate if excited by $\omega_1=2\omega_0(q)$. The resonance
does not require the actuation amplitude to exceed a finite threshold,
$P_{1{\rm c}}=0$; the growth rate simply increases linearly with $P_1$
(eq.~(\ref{Gamma_elastic})). This is due to the absence of damping
($K_1=0$).

\begin{figure}
  \centerline{\includegraphics[width=0.47\textwidth]{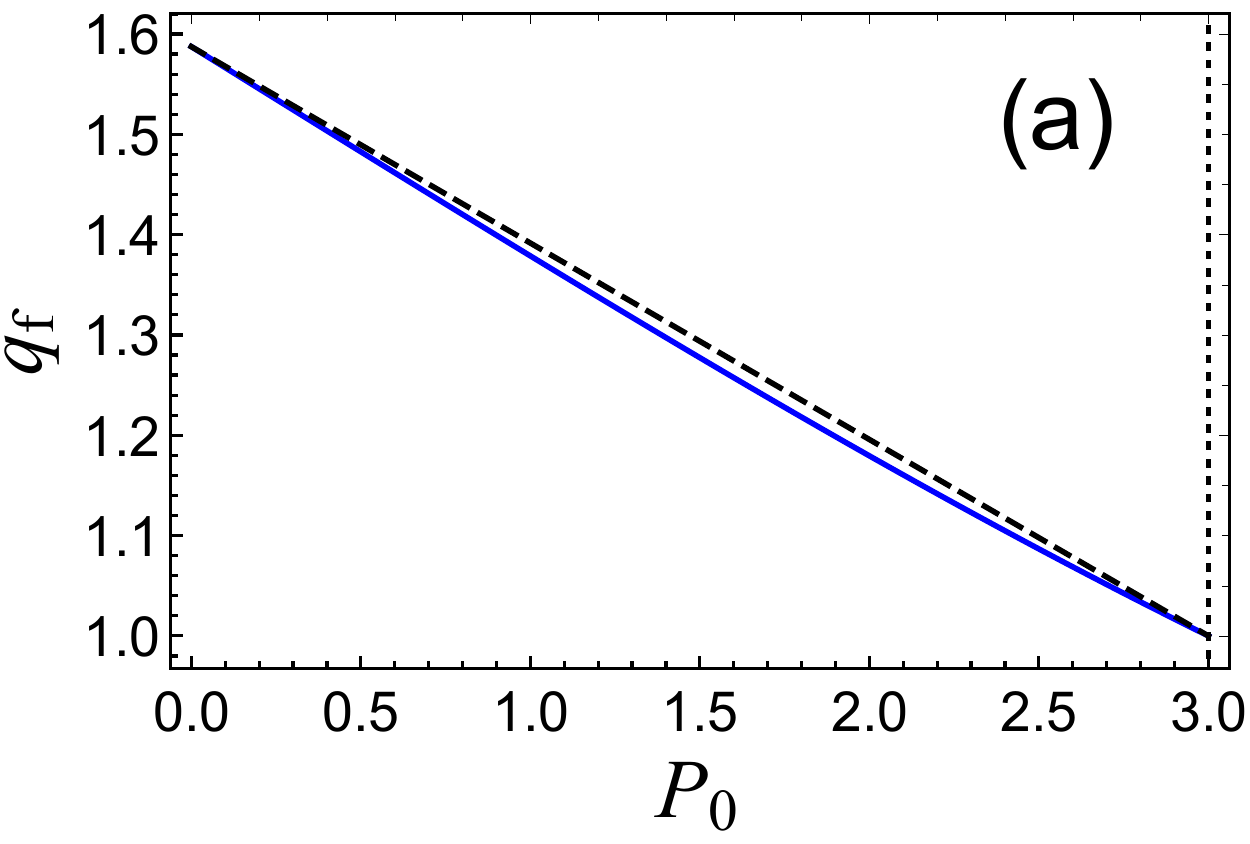}}
  \centerline{\includegraphics[width=0.47\textwidth]{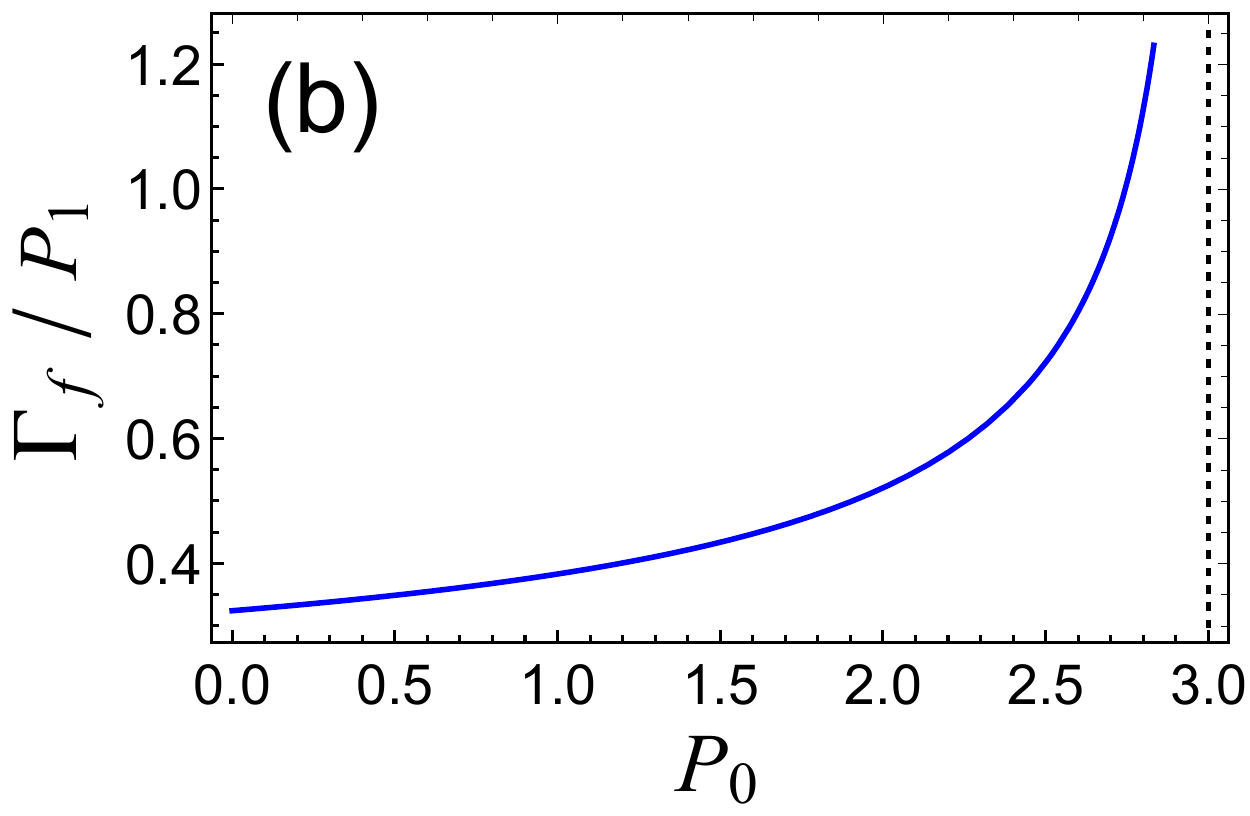}}
  \centerline{\includegraphics[width=0.47\textwidth]{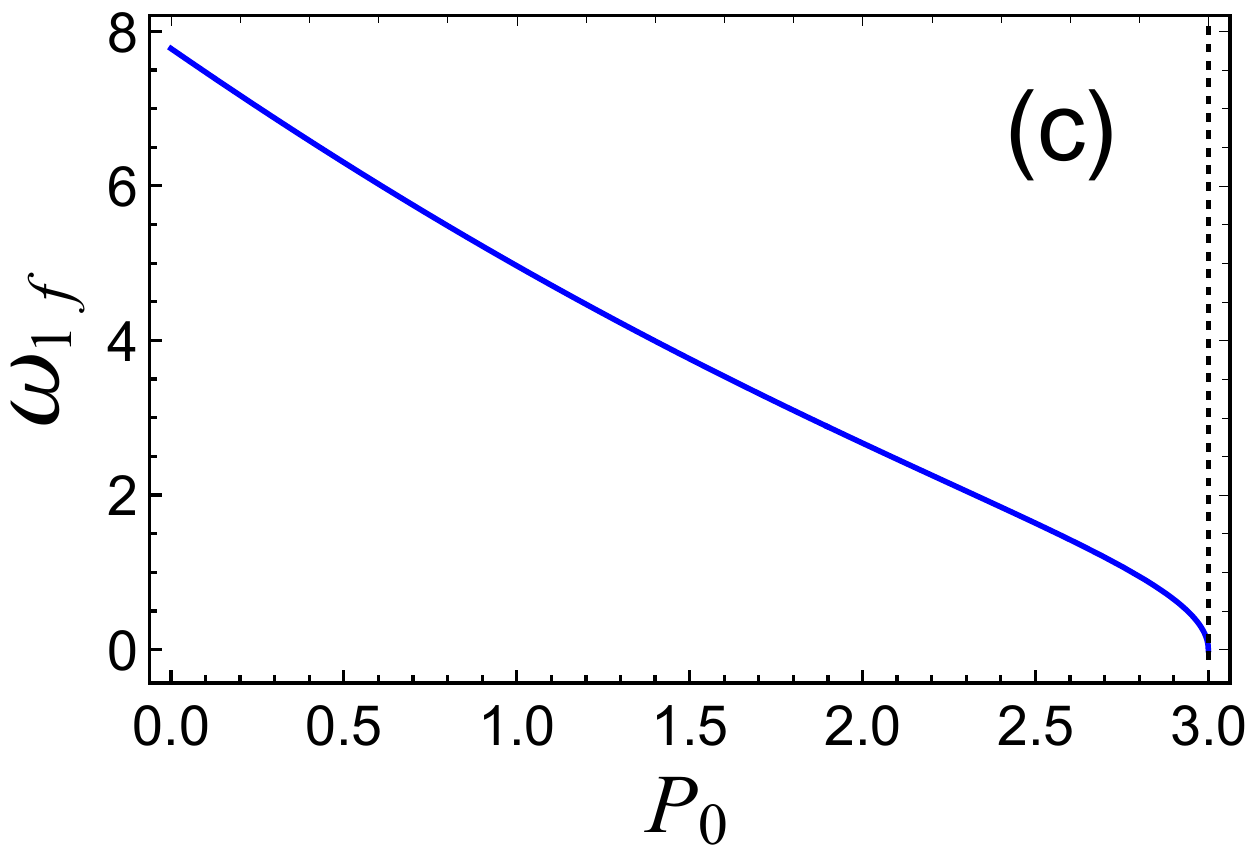}}
  \caption{Properties of the fastest-growing mode as a function of
    static pressure for an elastic substrate. (a) Wavenumber (solid
    line). The dashed line shows a linear interpolation between the
    analytically known wavenumbers $\qf(0)=2^{2/3}$ and
    $\qf(\Pzc)=1$. (b) Growth rate, diverging at $\Pzc$. For an
    elastic substrate, it is proportional to the actuation pressure
    $P_1$. (c) Actuation frequency required to excite the
    fastest-growing mode, vanishing at $\Pzc$. All parameters are
    normalized (see text).}
  \label{fig:fastest}
\end{figure}

Maximizing eq.~(\ref{Gamma_elastic}) gives the fastest-growing mode
$\qf(P_0)$ and its rate of amplitude growth $\Gammaf(P_0,P_1)$. These
functions are shown in fig.~\ref{fig:fastest}(a) and (b). Also shown,
in panel (c), is the actuation frequency $\omegaf(P_0)$ required to
excite the fastest-growing mode, as obtained from
eqs.~(\ref{zero_detuning}) and (\ref{omega0_elastic}). For $P_0=0$
(uncompressed sheet) we have $\qf=2^{2/3}\simeq 1.59$,
$\Gammaf/P_1=2^{-5/6}3^{-1/2}\simeq 0.324$, and
$\omegaf=2^{13/6}3^{1/2}\simeq 7.78$. Thus the fastest-growing
wavelength is smaller than that of the static wrinkles. As the static
pressure is increased, $\qf$ decreases (wavelength increases),
$\Gammaf/P_1$ increases, and $\omegaf$ decreases, until, at
$P_0=\Pzc=3$, the wavelength converges to the static one, $\omegaf$
vanishes, and $\Gammaf$ diverges.

\begin{figure}
  \centerline{
    \includegraphics[width=0.47\textwidth]{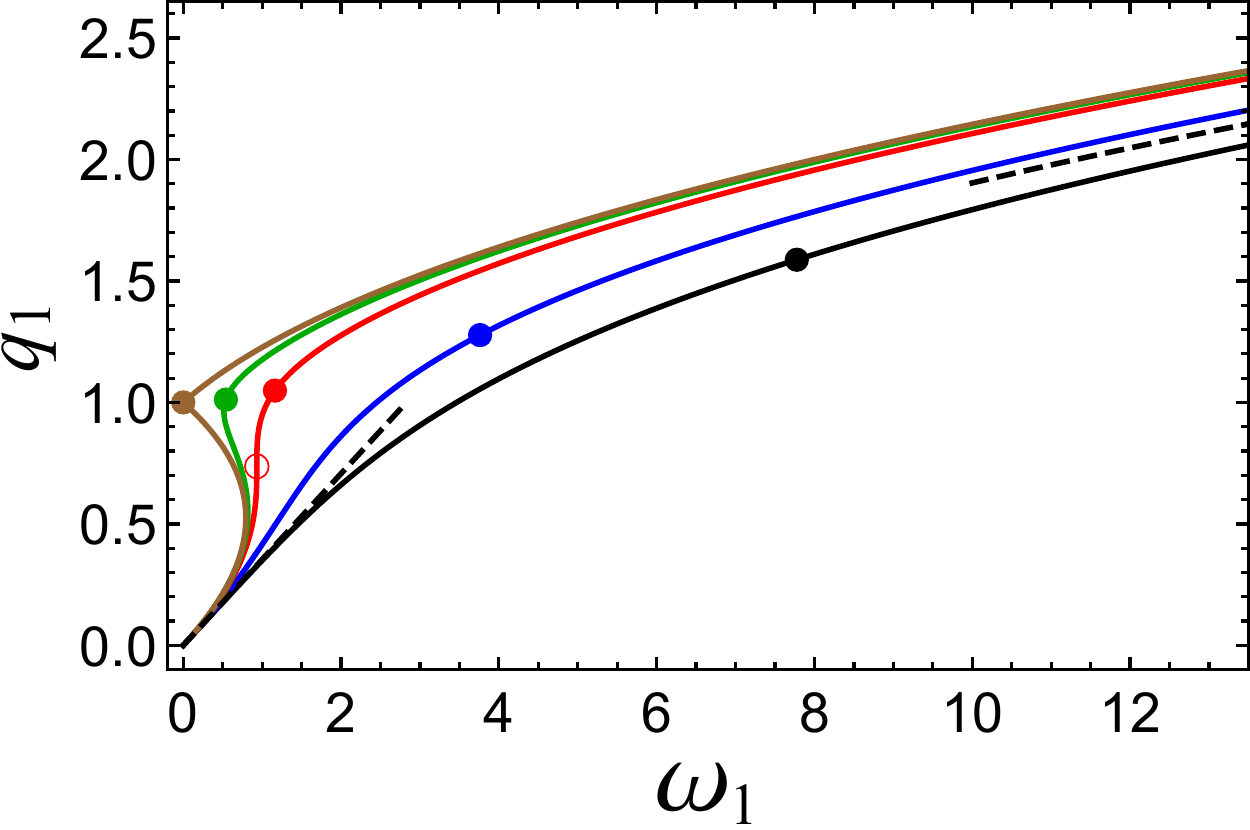}}
  \caption{Wrinkle wavenumber as a function of actuation frequency for
    an elastic substrate. Different curves correspond to different
    values of static pressure $P_0$ (from right to left): $0$, $1.5$,
    $P_0^*=20^{1/3}$, $2.93$, and $\Pzc=3$. Solid circles indicate the
    fastest-growing mode for the corresponding pressure. Dashed lines
    show the asymptotes given in eq.~(\ref{q1omega1asymp}). For
    $P_0>P_0^*$ there are three solutions for $q_1$, the largest of
    which growing the fastest, implying a discontinuous jump in the
    observed dominant wavenumber as $\omega_1$ is ramped up. The empty
    circle marks the bifurcation point. All parameters are normalized
    (see text).}
  \label{fig:q1w1el}
\end{figure}

The fastest-growing mode, however, is not the selected resonant mode.
The natural control parameters in experiment are the static pressure,
the actuation frequency, and the actuation amplitude.  Given $P_0$,
the choice of $\omega_1$ selects a dynamic wrinkle wavenumber,
$q_1(\omega_1,P_0)$, according to eqs.~(\ref{zero_detuning}) and
(\ref{omega0_elastic}). This wavenumber is not equal to $\qf$ in
general, and is independent of $P_1$. Figure~\ref{fig:q1w1el} shows
the selected wavenumber as a function of $\omega_1$ for several values
of $P_0$ between $0$ and $\Pzc$. The figure shows also the asymptotes
of $q_1$ for small and large $\omega_1$, which are both independent of
$P_0$,
\begin{equation}
  q_1(\omega_1,P_0) \simeq \left\{
  \begin{array}{ll} 2^{-3/2}\,\omega_1, & \omega_1\ll 1 \\
    (\omega_1/2)^{2/5}, \ \ \ \ \ \ & \omega_1 \gg 1. \end{array} \right. 
  \ \ \ \ \ \
\label{q1omega1asymp}
\end{equation}
The corresponding asymptotes for the rate of amplitude growth are
\begin{equation}
  \Gamma(\omega_1,P_0,P_1) \simeq \left\{
  \begin{array}{ll} (P_1/8)\,\omega_1, & \omega_1\ll 1 \\
    P_1 (16\,\omega_1)^{-1/5}, \ \ \ \ \ \ & \omega_1 \gg 1.
  \end{array} \right. 
  \ \ \ \ \ \
\end{equation}

The asymptotes in eq.~(\ref{q1omega1asymp}) confirm our earlier
statement, that $q_1$ is never much smaller than $\omega_1$, in
dimensionless terms. Switching for a moment back to dimensional
parameters, the two asymptotes become $q_1 \sim
(\rhom/G)^{1/2}\omega_1$ and $q_1 \sim (\rhom/B)^{1/5}\omega_1^{2/5}$,
revealing the different physical mechanisms in the two limits. At low
frequencies the restoring mechanism is the substrate's elasticity,
whereas at high frequencies it is the sheet's bending rigidity. This
crossover was anticipated in sect.~\ref{sec_scales}. Less expected is
the finding that the change between the two behaviors may be
discontinuous, as we shall see now.

At $P_0=P_0^*=(20)^{1/3}\simeq 2.71$ and
$\omega_1=\omega_1^*=2(2/5)^{5/6} \simeq 0.932$, the selected
wavenumber, which is at this point $q_1^*=(2/5)^{1/3} \simeq 0.737$,
bifurcates into three (fig.~\ref{fig:q1w1el}). The bifurcation entails
anomalous dynamics. At the bifurcation point we have $d\omega_0/d
q=0$, implying that an excitation with $P_0^*$ and $\omega_1^*$ at one
edge of the sheet will not propagate through the sheet. For
$P_0>P_0^*$ and $\omega_1<\omega_1^*$, we find from
eq.~(\ref{instability0}) that the largest of the three solutions for
$q_1(\omega_1,P_0)$ grows the fastest. Thus, for $P_0>P_0^*$, as the
excitation frequency $\omega_1$ is gradually increased from $0$, the
selected wrinkle wavenumber will undergo a discontinuous jump. For
increasingly larger static pressure $P_0$, the jump occurs at lower
and lower frequencies (see fig.~\ref{fig:q1w1el}), until, at
$P_0=\Pzc$, the system selects $q_1=\qc$ at zero frequency, as it
should. This is how the static-wrinkling limit is reproduced from the
dynamic one. Note that this entire behavior is independent of $P_1$;
hence, the discontinuous transition is present also for arbitrarily
weak actuation.

\begin{figure}
  \centerline{\includegraphics[width=0.47\textwidth]{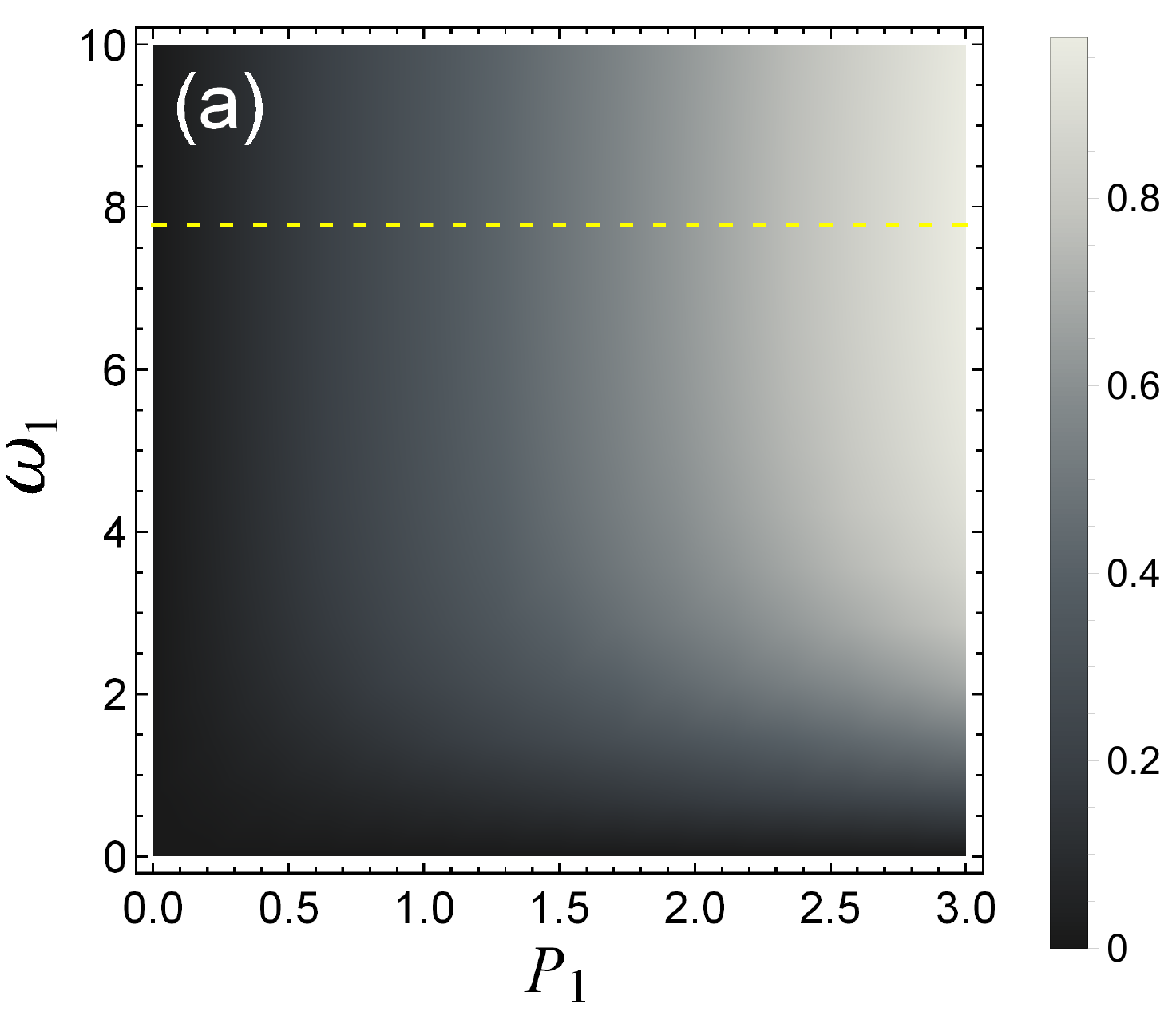}}
  \centerline{\includegraphics[width=0.47\textwidth]{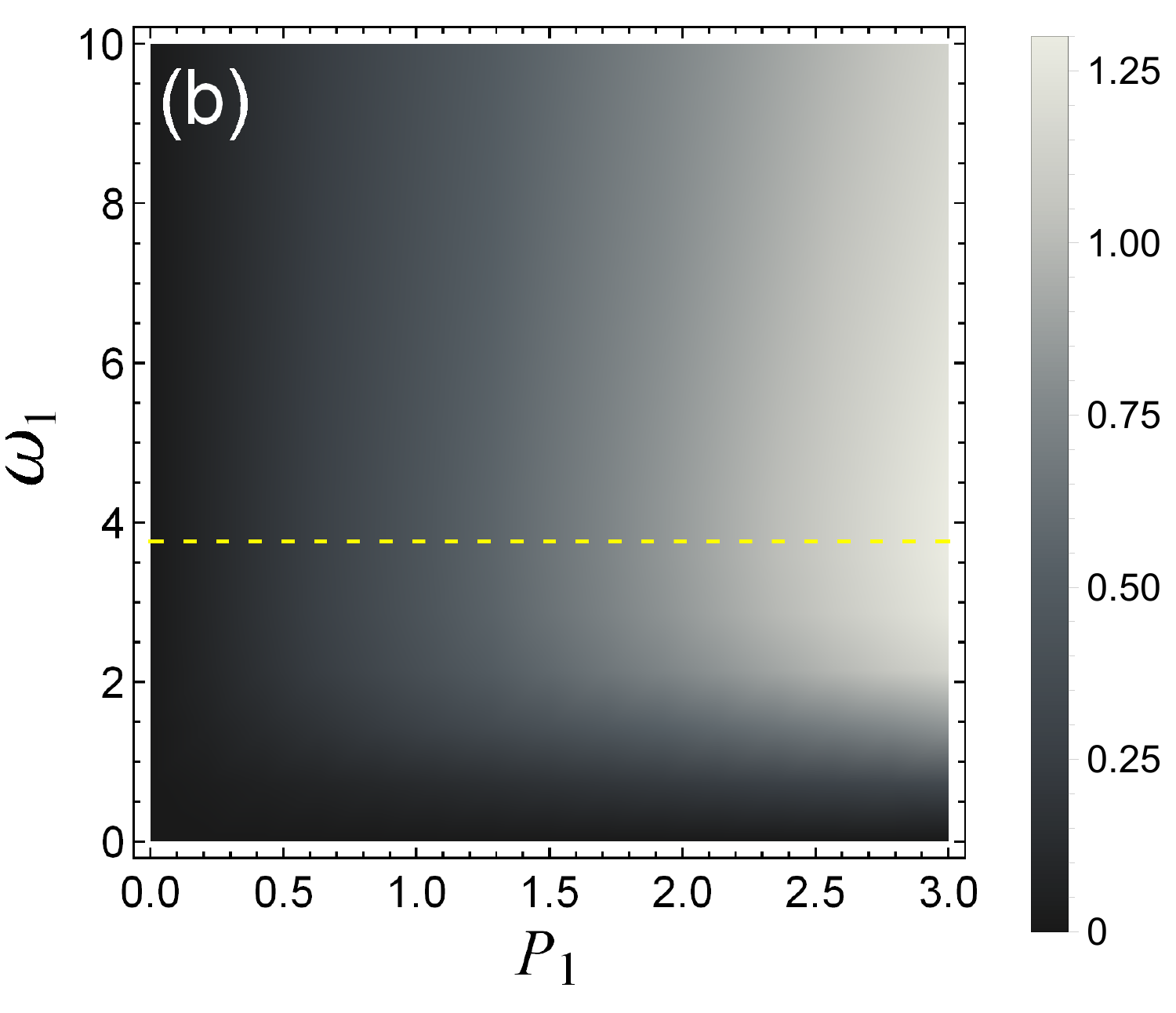}}
  \caption{Density plots of wrinkle growth rate as a function of
    excitation amplitude and frequency for an elastic substrate. The
    static pressure values are (a) $P_0=0$ and (b)
    $P_0=\Pzc/2=3/2$. The dashed lines show $\omegaf$, the
    excitation frequency that produces the fastest-growing mode (which
    for an elastic substrate is independent of $P_1$). All parameters
    are normalized (see text).}
  \label{fig:GP1w1el}
\end{figure}

Figure~\ref{fig:GP1w1el} presents 2D maps of the growth rate $\Gamma$
as a function of $P_1$ and $\omega_1$ for $P_0=0$ and $P_0=\Pzc/2$.

\subsection{Viscoelastic substrate}
\label{sec_ve}

For a viscoelastic medium the response is generalized by replacing $G$
with a frequency-dependent complex shear modulus
$\tilde{G}(\omega)$. (Recall that we have been assuming an
incompressible medium.) Applying the single-relaxation approximation
of eq.~(\ref{Kgeneral}), we generalize eq.~(\ref{K_elastic}) above to
\begin{equation}
  K_0 = 4G|q|,\ \ \ K_1 = 4\eta|q|,\ \ \ K_2 = \rhom/|q|,
\label{KthickVE}
\end{equation}
where $G={\rm Re}(\tilde{G})$ and $\eta={\rm Im}(\tilde{G})/\omega$
are the substrate's `store' modulus and viscosity, respectively
($\omega\eta$ is the `loss' modulus). We use the same units of energy,
length, and time as in sect.~\ref{sec_elastic}, making $B$,
$\hat{G}=2G$, and $\rhom$ all equal to unity. The viscosity $\eta$ is
measured then in units of $\rhom^{1/2} \hat{G}^{1/6} B^{1/3}$.

Substituting eq.~(\ref{KthickVE}) in eqs.~(\ref{resonanceparameters})
and (\ref{instability0}) while neglecting $\rho$, we obtain
\begin{eqnarray}
    \omega_0^2(q) &=& q^2(q^3 - 4\eta^2q^2 - P_0 q + 2),
\label{omega0_VE} \\
  \Gamma^2(q) &=& \left[ \frac{P_1^2}{4(q^3 - 4\eta^2q^2 - P_0q + 2)}
    - 16\eta^2 q^2 \right] q^2.
\label{Gamma_ve}
\end{eqnarray}

\begin{figure}
  \centerline{\includegraphics[width=0.47\textwidth]{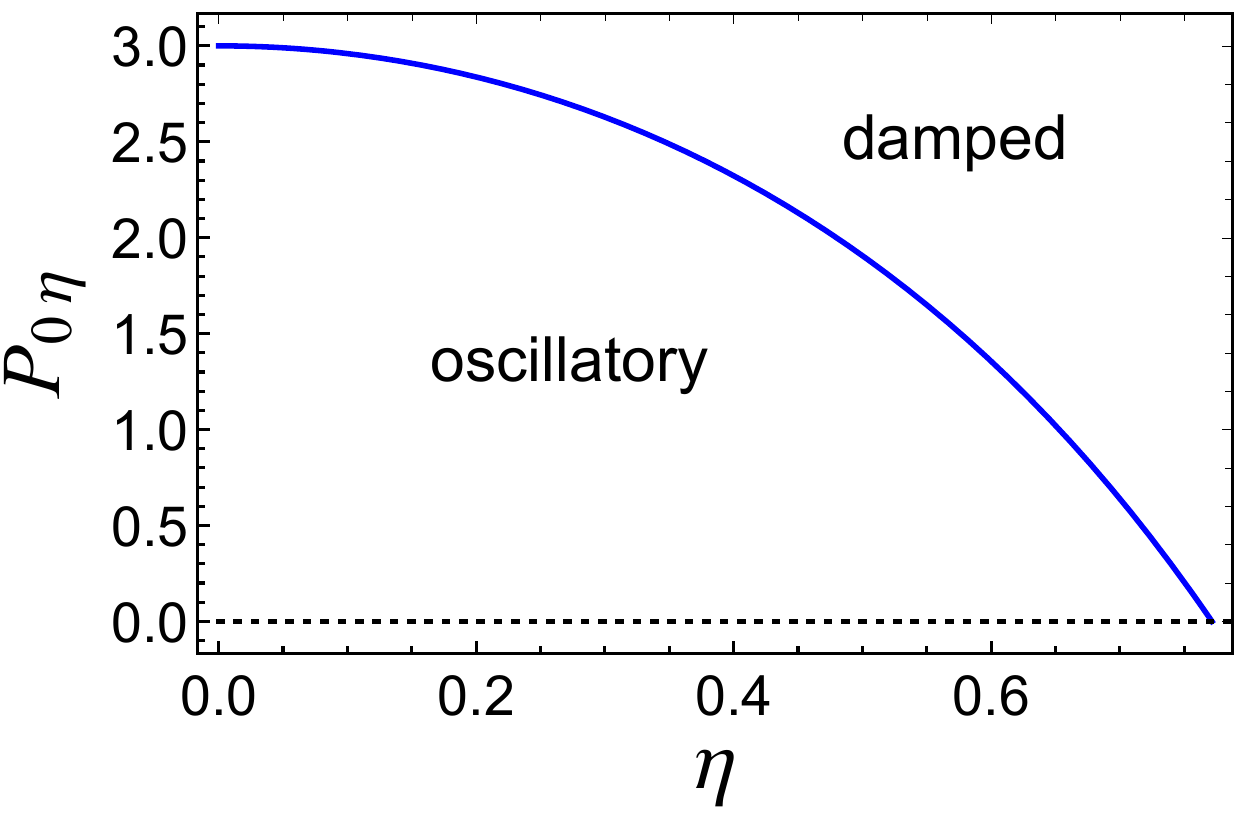}}
  \caption{Oscillatory and damped modes for a viscoelastic
    substrate. At static pressures smaller than $P_{0\eta}(\eta)$
    (solid curve), all modes are oscillatory; at larger pressures
    increasingly more modes are damped. For $\eta=0$ (elastic
    substrate) all modes are oscillatory for any $P_0<\Pzc=3$. All
    parameters are normalized (see text).}
  \label{fig:P0eta}
\end{figure}

The viscous component leads to several essential changes compared to
the elastic case. First, for a given $\eta$, $P_0$ has to be smaller
than some $P_{0\eta}(\eta)<\Pzc$ to have all modes oscillatory
($\omega_0^2>0$). The limiting function $P_{0\eta}(\eta)$ is shown in
fig.~\ref{fig:P0eta}. As one approaches the static wrinkling
transition, increasingly more modes become damped. For
$\eta>2^{-7/6}3^{1/2} \simeq 0.772$, $P_{0\eta}<0$, \ie there are
damped modes for any static pressure (unless we `strengthen the
springs' by stretching the sheet with $P_0<0$).

As in the elastic case, fixing $\omega_1=2\omega_0$ selects a mode,
$q_1(\omega_1,P_0,\eta)$, which does not depend on $P_1$ (see
eq.~(\ref{omega0_VE})). Unlike the elastic case, the fastest-growing
mode $\qf(P_0,P_1,\eta)$, obtained by maximizing $\Gamma$ of
eq.~(\ref{Gamma_ve}), does depend on $P_1$. Hence, the fastest-growing
mode does not belong in general to the set of selected wavenumbers;
one should tune $P_1$ together with $\omega_1$ to get $q_1=\qf$ (see
fig.~\ref{fig:GP1w1ve} below). Figure~\ref{fig:q1w1ve}(a) shows the
selected wavenumber as a function of $\omega_1$ for several values of
$P_0$ between $0$ and $\Pzc$. The asymptotes for small and large
$\omega_1$ remain as in eq.~(\ref{q1omega1asymp}). Also here, the
solutions bifurcate above a certain static pressure $P_0^*$, implying
a discontinuous jump in the dominant wrinkle wavenumber as $\omega_1$
is increased. The bifurcation point depends now on $\eta$. (See the
Supplementary Material \cite{suppl} for the functional dependence.)
Figure~\ref{fig:q1w1ve}(b) shows the decrease of $P_0^*$ with
$\eta$. For sufficiently high viscosity, $\eta > 3^{1/2}5^{1/3}/4
\simeq 0.740$, the change of wavenumber with frequency is
discontinuous for any $P_0$. Finally, for $P_0>P_{0\eta}>P_0^*$ a band
of modes becomes damped (with imaginary $\omega_0$) as manifested by
the leftmost curve in fig.~\ref{fig:q1w1ve}(a).

\begin{figure}
  \centerline{
    \includegraphics[width=0.47\textwidth]{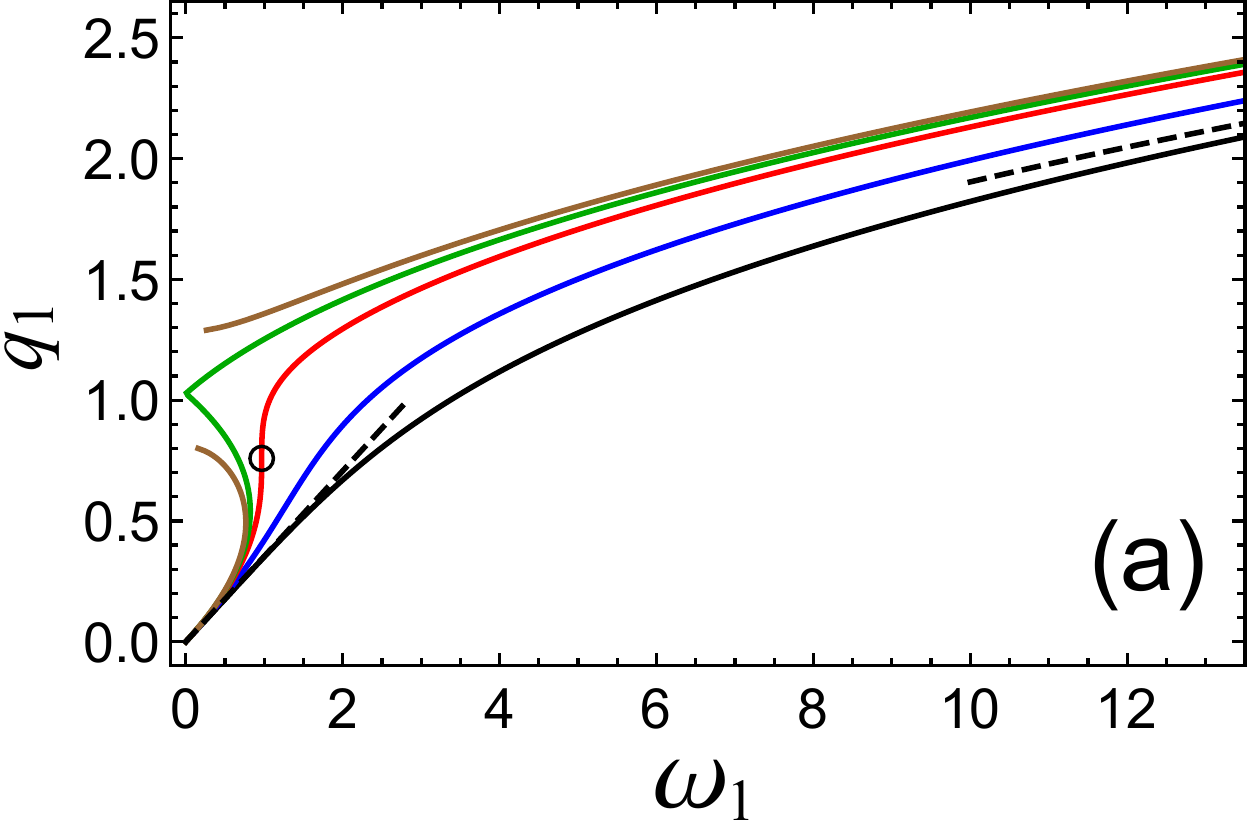}}
  \centerline{
    \includegraphics[width=0.47\textwidth]{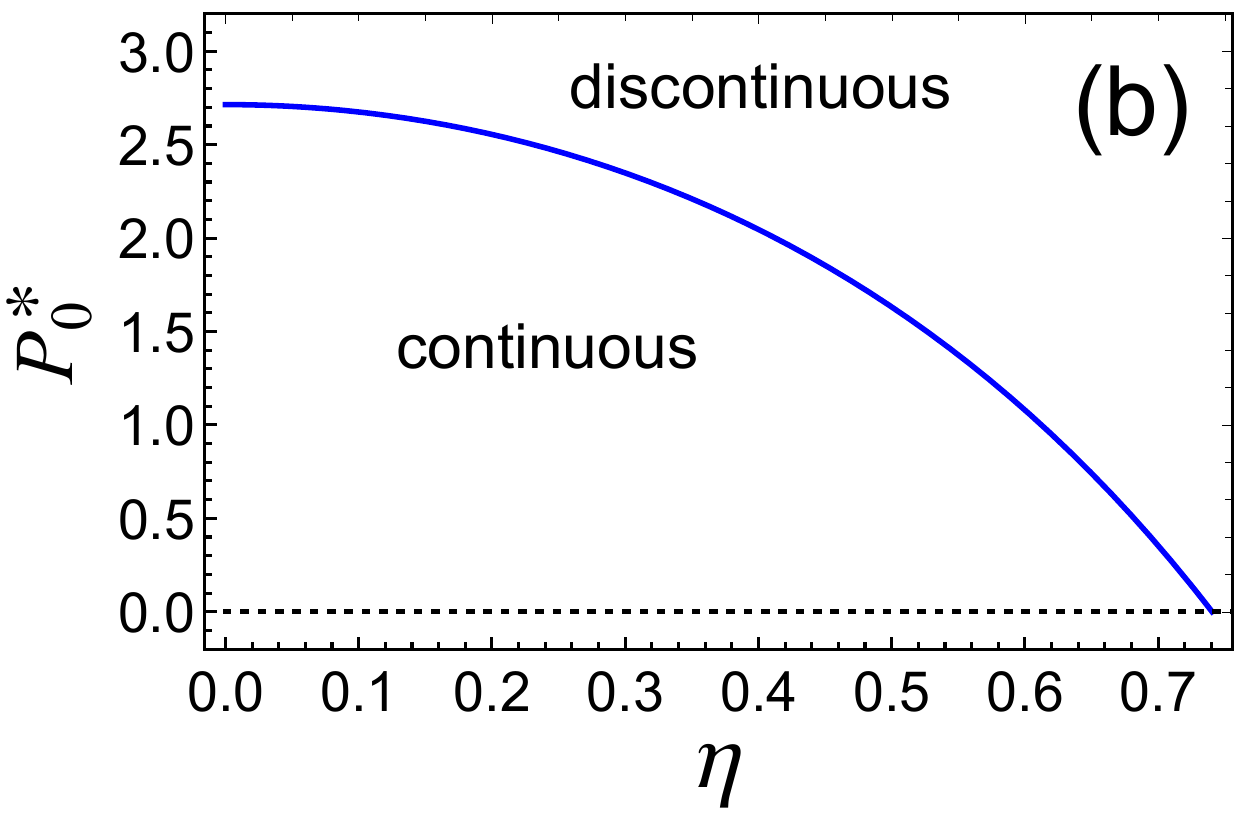}}  
  \caption{Change of wrinkle wavenumber with actuation frequency for a
    viscoelastic substrate. (a) Wavenumber as a function of frequency
    for a given viscosity, $\eta=0.2$. Different curves correspond to
    different values of static pressure $P_0$ (from right to left):
    $0$, $\Pzc/2=3/2$, $P_0^*= 2.55$, $P_{0\eta}=2.84$, and
    $\Pzc=3$. Dashed lines show the asymptotes given in
    eq.~(\ref{q1omega1asymp}). For $P_0>P_0^*$ there are three
    solutions for $q_1$, the largest of the three growing the fastest,
    implying a discontinuous jump in the observed dominant wavenumber
    as $\omega_1$ is ramped up. The empty circle marks the bifurcation
    point. For $P_0>P_{0\eta}$ a band of modes are damped (leftmost,
    brown curve). (b) Decrease of the bifurcation pressure with
    viscosity. All parameters are normalized (see text).}
  \label{fig:q1w1ve}
\end{figure}

\begin{figure}
  \centerline{
    \includegraphics[width=0.47\textwidth]{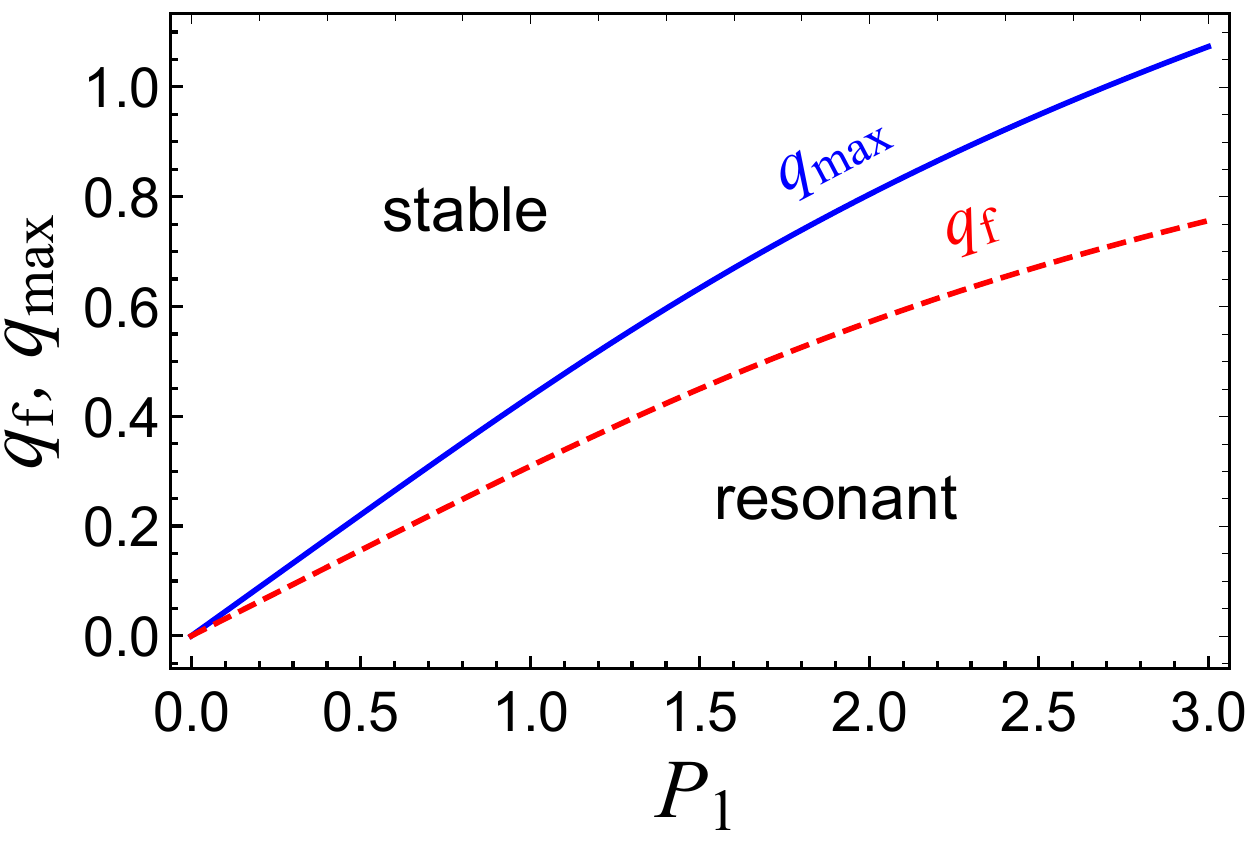}}
  \caption{Unstable (resonant) band as a function of actuation
    amplitude for an uncompressed sheet ($P_0=0$) on a viscoelastic
    substrate. The viscosity is $\eta=0.2$. For $q$ larger than the
    cutoff $\qmax$ (solid line), the modes are stable (of finite
    amplitude). Also shown is the fastest-growing mode $\qf$ (dashed
    line). All parameters are normalized (see text).}
  \label{fig:qfqmax}
\end{figure}

Another important change brought about by viscosity is that the
oscillatory modes do not resonate for every value of $P_1$ and
$\omega_1$. The expression for the squared growth rate in
eq.~(\ref{Gamma_ve}) has the asymptotes $(P_1^2/8)q^2$ and $-16\eta^2
q^4$, respectively, at small and large $q$. Thus, for any finite $P_1$
there are small-$q$ resonant modes, but the unstable band has a
cutoff at some $\qmax(P_0,P_1,\eta)$. The reason why resonance should
require small wavenumber lies in the dependence of inertia on $q$
(cf.\ $K_2$ in eq.~(\ref{KthickVE})). The larger the wavelength, the
thicker the layer of substrate which moves with the sheet, and the
larger its inertia. Figure~\ref{fig:qfqmax} shows the dependence of
the cutoff $\qmax$, along with the fastest-growing mode $\qf$, on
$P_1$ for an uncompressed sheet ($P_0=0$) and a given viscosity.
Equation~(\ref{Gamma_ve}) can be rewritten as
\begin{equation}
  \Gamma^2 = \left( \frac{P_1^2}{\omega_1^2} - 16\eta^2 \right) q^4.
\end{equation}
Hence, resonance requires crossing a threshold of actuation amplitude,
which is linear in the actuation frequency,
\begin{equation}
  P_1 > P_{1{\rm c}} = 4\eta \omega_1.
\label{P1c}
\end{equation}
This is a consequence of the viscous damping. The smaller the
frequency, the weaker the actuation needed to overcome the
damping. The reason, once again, is that a larger mass of substrate is
involved in the motion for small \\ wavenumber (low frequency).

Figure~\ref{fig:GP1w1ve} shows a 2D map of the growth rate $\Gamma$ as
a function of the excitation parameters $P_1$ and $\omega_1$ for an
uncompressed sheet ($P_0=0$). Unlike the elastic-substrate case
(fig.~\ref{fig:GP1w1el}), here the resonant region is bounded.

\begin{figure}
  \centerline{\includegraphics[width=0.47\textwidth]{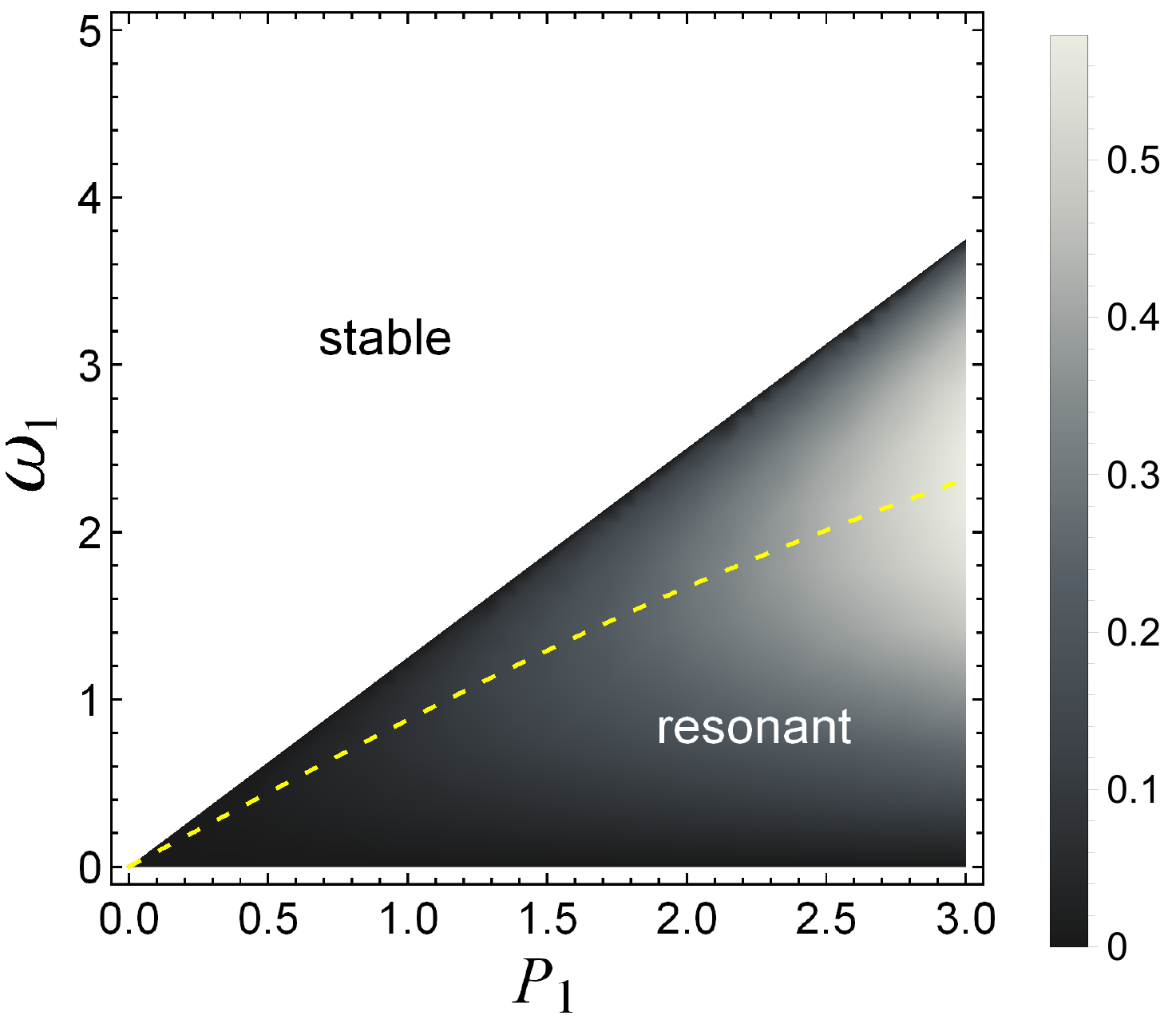}}
  \caption{Density plot of wrinkle growth rate as a function of
    actuation amplitude and frequency for an uncompressed sheet
    ($P_0=0$) on a viscoelastic substrate. The viscosity is
    $\eta=0.2$. The resonant band is bounded by a minimum amplitude
    proportional to $\omega_1$. The dashed line shows the excitation
    frequency that produces the fastest-growing mode for each
    actuation amplitude. All parameters are normalized (see text).}
  \label{fig:GP1w1ve}
\end{figure}

\section{Discussion}
\label{sec_discuss}

\subsection{Summary of experimental predictions}
\label{sec_experiment}

Let us summarize the results which seem most relevant experimentally,
and give them in dimensional form. As in the analysis above, we assume
that the inertia is governed by the substrate. A similar discussion
for the case of sheet-dominated inertia is given in the Supplementary
Material \cite{suppl}.

In the case of an elastic substrate, one can first compress the sheet
until static wrinkling occurs. The measured critical pressure and
static wrinkle wavenumber are related to the bending modulus of the
sheet and the elastic modulus of the substrate as
\begin{equation}
  \Pzc = 3 B^{1/3} G^{2/3},\ \ \
  \qc = (G/B)^{1/3},
\label{Pcqcdimensional}
\end{equation}
with known corrections for finite compressibility
\cite{Groenewold2001}. This allows a measurement of $B$ and $G$.

For a finite $P_0<\Pzc$, and ramping up the actuation frequency
$\omega_1$ from zero, dynamic wrinkles should form for any actuation
amplitude. At low frequency ($\omega_1\ll\omegam
\sim(G/\rhom)^{1/2}/\lambdac$), the wrinkle wavenumber $q_1$ increases
linearly with $\omega_1$,
\begin{subequations}
\begin{equation}
  q_1 \simeq (1/4) (\rhom/G)^{1/2}\, \omega_1,
 \label{q1w1dimensional1}
\end{equation}
which is essentially the relation for Rayleigh surface waves
\cite{LLelasticityRayleigh}.  At high frequencies
($\omega_1\gg\omegam$) the wavenumber increases as $\omega_1^{2/5}$,
\begin{equation}
  q_1 \simeq 0.758 (\rhom/B)^{1/5}\, \omega_1^{2/5}.
 \label{q1w1dimensional2}
\end{equation}
\label{q1w1dimensional}
\end{subequations}
Concerning the dependence on sheet thickness, at low frequencies the
wrinkle wavenumber is independent of $h$, and at high frequencies it
decreases with $h$ as $\sim h^{-3/5}$. These very different
dependencies are related to the different restoring mechanisms in the
two regimes. At low frequency the substrate's elasticity dominates,
and the resulting Rayleigh waves are independent of the sheet. At high
frequency the dominant force comes from the sheet's bending rigidity,
which depends on thickness. The two dependencies are to be compared
with that of the static wrinkles, where $\qc \sim h^{-1}$; see
eq.~(\ref{Pcqcdimensional}).

Depending on the value of $P_0$, two distinct behaviors are expected
as $\omega_1$ is increased. At small pressures, $P_0<P_0^*$, the
selected wrinkle wavelength decreases continuously with
$\omega_1$. For larger pressures, $P_0^*<P_0<\Pzc$, a discontinuous
drop in the dominant wavelength is expected as a function of
$\omega_1$. The transition occurs at
\begin{equation}
  P_0^* = 0.903\Pzc.
\label{P0sdimensional}
\end{equation}
The transition in the wrinkle wavelength is a particularly distinctive
prediction. We discuss its validity further in
sect.~\ref{sec_extension} below.

The behavior in the case of a viscoelastic substrate is qualitatively
different. Thus it might be used to obtain information on the
viscoelastic properties of the supporting medium. The present theory
is restricted, however, to the simple case where the viscoelastic
response is described sufficiently well by a single relaxation time,
$\tau=\eta/G$, \ie the complex modulus is given by $\tilde{G} = G +
i\omega\eta$; cf.\ sects.~\ref{sec_eqs} and \ref{sec_ve}. The static
measurement of $\Pzc$ and $\qc$ are as in the elastic case above.

To see dynamic wrinkles on a viscoelastic substrate one needs an
excitation with pressure amplitude that exceeds a threshold $P_{1{\rm
    c}}$. The threshold depends linearly on the excitation frequency,
\begin{equation}
  P_{1{\rm c}} = 3.17 \eta (B/G)^{1/3}\, \omega_1=
  3.17 \eta \lambdac\, \omega_1.
\end{equation}
where $\lambdac$ is the static wrinkle wavelength. Thus the threshold
of resonance may be used as a probe of the viscous component
$\eta$. As in the elastic case, at low and high excitation frequencies
the asymptotic dependence of the dynamic wrinkle wavenumber $q_1$ on
$\omega_1$ is given in eqs.~(\ref{q1w1dimensional}). The remark
concerning the dependence on sheet thickness in the elastic case holds
here as well.

In the viscoelastic case, too, the value of $P_0$ separates the
behaviors when ramping up $\omega_1$ into two cases: a continuous
decrease of wavelength for low pressure and a discontinuous one at
high pressure. The transition pressure $P_0^*$ decreases with
viscosity (see fig.~\ref{fig:q1w1ve}(b) and the Supplementary Material
\cite{suppl}), providing another probe of $\eta$.

To get a feeling for the relevant scales, we consider a specific
system, motivated by the experimental system of
ref.~\cite{Pocivavsek2018}. It is made of a $1$-mm-thick stiffer
elastomeric sheet ($\Es \sim 10^6$~Pa), supported on a softer
elastomeric medium ($\Em\sim 10^4$~Pa). These properties fit also a
layer of skin covering a muscle tissue. The sheet's bending modulus is
$B\sim 10^{-4}$~J. The resulting static wrinkle wavenumber
(eq.~(\ref{Pcqcdimensional})) is $\qc \sim 1$~mm$^{-1}$. (This is at
the edge of the theory's validity, which requires $qh\ll 1$; thus the
following should be regarded only as qualitative orders of magnitude.)
To excite dynamic wrinkles of a similar wavenumber we need, according
to eq.~(\ref{q1w1dimensional}), an excitation frequency of order
$\omega_1\sim 10^4$~s$^{-1}$. (We have taken $\rhom\sim
10^3$~kg/m$^3$.) This is close to the relevant lower frequency bound
obtained in sect.~\ref{sec_intro}. As already noted there, such
frequencies are probably too high to be produced naturally but readily
attainable in experiments.

To observe the viscous effects described in sect.~\ref{sec_ve}, we
need a normalized $\eta$ of order $1$. In dimensional terms it
implies, for the example above, $\eta \gtrsim 1$--$10$~Pa\,s (\ie
$10^3$--$10^4$ times the viscosity of water). This is in line with the
estimate in sect.~\ref{sec_scales}.

\subsection{Comparison with other dynamic wrinkling scenarios}

As mentioned in sect.~\ref{sec_intro}, several works have addressed
the formation of dynamic wrinkles in thin sheets upon time-varying
external forcing, whose source may be, for example, the impact of a
rigid object, or an abrupt change of pressure or confinement
\cite{Vermorel2009,Vandenberghe2016,Box2019,Ghanem2019,Kodio2017,Chopin2017,Box2020}. The
main feature that sets the system addressed here apart is the
periodic, single-frequency external forcing. Within our linear theory,
it implies the selection of a single, constant wrinkle wavelength. For
the non-periodic forcing in the other scenarios, a time-increasing
(coarsening) wavelength has been observed (\eg
ref.~\cite{Box2019}). Unlike the periodically excited system, the
other systems eventually approach equilibrium whereby the dynamic
wavelength must tend toward its static value.

In addition, parametric resonance has a different mode-selection
mechanism. The mode $q_1$ is selected to match the actuation frequency
(such that $2\omega_0(q_1)=\omega_1$). Thus it is not equal to the
fastest-growing mode $\qf$, which is the selected mode in the other
scenarios. One consequence concerns the dependence of the selected
mode on inertia.  In the absence of damping, the mode which maximizes
the growth rate in our system is independent of $\rhom$. (See
eqs.~(\ref{resonanceparameters}) and (\ref{instability0}) in which,
for $K_1=0$, the mass density enters only in a prefactor.)  Similarly,
the selected pattern in other dynamic-buckling systems was found to be
independent of inertia (\eg ref.~\cite{Box2020}). In the resonant
system the selected mode $q_1$ does not maximize the growth rate and
thus depends on $\rhom$.

Nevertheless, there is a qualitative relation with the time dependence
of the selected mode in a supported sheet under impact
\cite{Box2019}. In ref.~\cite{Box2019} the selected wavelength was
found to increase with time according to \\ $\lambda(t) \sim
(B/\rhom)^{1/5}\,t^{2/5}$. This scaling is in line with our
$q_1(\omega_1)$ relation in the high-frequency limit,
eq.~(\ref{q1w1dimensional2}). It arises in both cases from the
interplay of sheet bending and substrate inertia (see
sect.~\ref{sec_scales}). Consistently with this limit, the wavelength
values measured in the impact experiments were much smaller than
$\lambdac$. With a compressed sheet on a (visco)elastic substrate, the
impact behavior at longer times (corresponding to our
low-frequency-large-wavelength limit, eq.~(\ref{q1w1dimensional1}))
might reveal an instability or a two-wave pattern similar to the one
predicted above for $q_1(\omega_1)$.

\subsection{Model extensions}
\label{sec_extension}

We have assumed above that the inertia is governed by the
substrate. As estimated in sect.~\ref{sec_intro}, this is valid when
the wavelength is much larger than the sheet thickness. When the two
are not scale-separated, the sheet's inertia may be important. (In
fact, this may be the case in the numerical example given above.) The
physical difference between the two limits is the fact that the
effective 2D mass responsible for inertia in the substrate case
depends on wavelength (cf.\ $K_2$ of eq.~(\ref{K_elastic_both})),
whereas for the sheet it is a constant. The combination of inertial
effects from both substrate and sheet can be treated within our
theory. One should return to the equations of motion,
sect.~\ref{sec_eqs}, and consider the full inertial terms with $\rhos
h+K_2(q)$ instead of just $K_2$. The algebra is more cumbersome but
can be treated numerically.

The opposite limit, of sheet-dominated inertia, is presented in the
Supplementary Material \cite{suppl}. Although this limit is of less
practical relevance, it is instructive to see the qualitative changes
brought about by the sheet's mass. These are as follows. (a) The
fastest growing wavelength for an uncompressed sheet ($P_0=0$) is
arbitrarily small (whereas with substrate inertia it is
$\sim\lambdac$; see fig.~\ref{fig:fastest}). Thus a finite static
pressure is required to get finite-size dynamic wrinkles. (b) The
selected wavenumber scales differently with actuation frequency, as
$\omega_1^2$ and $\omega_1^{1/2}$ at low and high frequency,
respectively. (Compare to eq.~(\ref{q1omega1asymp}).) (c) As a result
of (b), the dependence of wrinkle wavenumber on sheet thickness is
different\,---\,increasing as $h$ and decreasing as $h^{-1/2}$ for
large and small $\omega_1$, respectively (compared to $h^0$ and
$h^{-3/5}$ with substrate inertia). Overall, however, the qualitative
behaviors are quite similar. In particular, the phenomenon of
continuous vs.\ discontinuous change of selected wavelength with
frequency exists in both limits.

We have used the large-frequency asymptotic form of the substrate's
kernel, eq.~(\ref{K_elastic}). The small-frequency asymptote is the
same up to numerical prefactors (see eq.~(\ref{K_elastic_both})), and
will lead to the same results. A more complete theory should consider
the full kernel, eq.~(\ref{KLamb}). This would require a more
complicated numerical analysis. One might be worried that our central
prediction, concerning the continuous vs.\ discontinuous behavior of
$q_1$ as a function of $\omega_1$, is an artifact of the asymptotic
kernel, as the phenomenon occurs at $q_1\lambdac \sim 1$ (see
figs.~\ref{fig:q1w1el} and \ref{fig:q1w1ve}). This is most probably
not the case. The transition is a result of the function
$\omega_0^2(P_0,q)$ becoming non-convex at sufficiently high
pressure. It is a generic property required to obtain the static
wrinkling transition, $\omega_0^2(\Pzc,\qc)=0$, at a finite wavenumber
$\qc$. Indeed, the case of sheet-dominated inertia \cite{suppl}, where
the much simpler kernel of a static elastic substrate is fully
treated, exhibits the same behavior.

The theory presented here is linear. As a result, it provides the
properties of the instability but not the ultimate form of the sheet's
dynamic deformation. Whether the deformation saturates to periodic
wrinkles of finite height, develops multi-wavelength wrinkles
\cite{Brau2011}, or localizes into deeper features (folds)
\cite{Brau2013,Box2019}, should be checked in a future nonlinear
theory or simulation.

We have assumed a semi-infinite substrate. Over length scales
comparable and larger than the substrate thickness the results will be
modified. In the opposite limit, of a thin substrate compared to the
wrinkle wavelength, the effect of the medium will turn into that of a
Winkler foundation \cite{Dillard2018}, \ie strongly localized ($\tK$
independent of $q$).

Another simplification employed here is the assumption of a single
relaxation time for the viscoelastic medium. Actual viscoelastic
media, particularly biological ones, have a much richer frequency
dependence, which will affect the response to the parametric
excitation. Conversely, parametric resonance may be used to tap into
the medium's rich temporal response based on an extended theory.

Besides relaxation times, complex media have also characteristic
lengths which affect their response
\cite{SonnSegev2014,Grosberg2016}. The present theory describes a way
to sample various length scales (wavenumbers) by sweeping the
parametric-excitation frequency. Recently we have derived the solution
to the Boussinesq problem for a viscoelastic structured medium,
accounting for its intrinsic correlation length
\cite{BarHaim2020}. Similar to the derivations in
sects.~\ref{sec_elastic} and \ref{sec_ve}, these results (once
extended to include inertia) may be used to address the parametric
excitation of a sheet supported on such a structured medium.

\begin{acknowledgement}
  Helpful discussions with Benny Davidovitch are gratefully
  acknowledged.
\end{acknowledgement}


\begin{thebibliography}{99}

\bibitem{Cerda2003} E. Cerda, L. Mahadevan, Geometry and physics of
  wrinkling, Phys. Rev. Lett. {\bf 90}, 074302 (2003).
  
\bibitem{Genzer2006} J. Genzer, J. Groenewold, Soft matter with hard
  skin: From skin wrinkles to templating and material characterization,
  Soft Matter {\bf 2}, 310--323 (2006).

\bibitem{Davidovitch2011} B. Davidovitch, R. D. Schroll, D. Vella,
  M. Adda-Bedia, E. Cerda,
  Prototypical model for tensional wrinkling in thin sheets,
  Proc. Natl. Acad. Sci. USA {\bf 108}, 18227--18232 (2011).

\bibitem{Pocivavsek2018} L. Pocivavsek, J. Pugar, R. O'Dea, S.-H. Ye,
  W. Wagner, E. Tzeng, S. Velankar, E. Cerda, Topography-driven
  surface renewal, Nat. Phys. {\bf 14}, 948--953 (2018).

\bibitem{Pocivavsek2019} L. Pocivavsek, S.-H. Yea, J. Pugar,
  E. Tzeng, E. Cerda, S. Velankar, W. R. Wagnera, Active wrinkles to
  drive self-cleaning: A strategy for anti-thrombotic surfaces for
  vascular grafts, Biomat. {\bf 192}, 226--234 (2019).

\bibitem{Nath2020} N. N. Nath, L. Pocivavsek, J. A. Pugar, Y. Gao,
  K. Salem, N. Pitre, R. McEnaney, S. Velankar, E. Tzeng, Dynamic
  luminal topography: A potential strategy to prevent vascular graft
  thrombosis, Front. Bioeng. Biotech. {\bf 8}, 573400 (2020).
  
\bibitem{Lin2020} G. Lin, W. Sun, P. Chen, Topography-driven
  delamination of thin patch adhered to wrinkling surface,
  Int. J. Mech. Sci. {\bf 178}, 105622 (2020).

\bibitem{Wen2020} X. Wen, S. Sun, P. Wu, Dynamic wrinkling of a
  hydrogel-elastomer hybrid microtube enables blood vessel-like
  hydraulic pressure sensing and flow regulation, Mater. Horiz. {\bf
    7}, 2150 (2020).
  

\bibitem{Vella2009} D. Vella, J. Bico, A. Boudaoud, B. Roman, P. M. Reis,
  The macroscopic delamination of thin films from elastic substrates,
  Proc. Natl. Acad. Sci. USA {\bf 106}, 10901--10906 (2009).

\bibitem{Mei2011} H. Mei, C. M. Landis, R. Huang, Concomitant
  wrinkling and buckle-delamination of elastic thin films on compliant
  substrates, Mech. Mater. {\bf 43}, 627--642 (2011).

\bibitem{Hohfeld2015} E. Hohfeld, B. Davidovitch, Sheet on a
  deformable sphere: Wrinkle patterns suppress curvature-induced
  delamination, Phys. Rev. E {\bf 91}, 012407 (2015).

\bibitem{Oshri2018} O. Oshri, Y. Liu, J. Aizenberg, A. C. Balazs,
  Delamination of a thin sheet from a soft adhesive Winkler substrate,
  Phys. Rev. E {\bf 97}, 062803 (2018).
  
\bibitem{Oshri2020} O. Oshri, Delamination of open cylindrical shells
  from soft and adhesive Winkler's foundation, Phys. Rev. E {\bf 102},
  033001 (2020).

\bibitem{Bixler2012} G. D. Bixler, B. Bhushan,
  Biofouling: Lessons from nature, Phil. Trans. R. Soc. A {\bf 370},
  2381--2417 (2012).

\bibitem{Sridhar2001} N. Sridhar, D. J. Srolovitz, Z. Suo, Kinetics of
  buckling of a compressed film on a viscous substrate,
  Appl. Phys. Lett. {\bf 78}, 2482--2484 (2001).

\bibitem{Huang2002} R. Huang, Z. Suo, Wrinkling of a compressed
  elastic film on a viscous layer, J. Appl. Phys. {\bf 91}, 1135--1142
  (2002).

\bibitem{Huang2005} R. Huang, Kinetic wrinkling of an elastic film on
  a viscoelastic substrate, J. Mech. Phys. Solids {\bf 53}, 63--89
  (2005).

\bibitem{Vermorel2009} R. Vermorel, N. Vandenberghe, E. Villermaux,
  Impacts on thin elastic sheets, 
  Proc. Roy. Soc. A {\bf 465}, 823--842 (2009).


\bibitem{Vandenberghe2016} N. Vandenberghe, L. Duchemin,
  Impact on floating membranes,
  Phys. Rev. E {\bf 93}, 052801 (2016). 

\bibitem{Box2019} F. Box, D. O'Kiely, O. Kodio,
  M. Inizan, A. A. Castrej\'on-Pita, D. Vella,
  Dynamics of wrinkling in ultrathin elastic sheets,
  Proc. Natl. Acad. Soc. USA {\bf 116}, 20875--20880 (2019). 

\bibitem{Ghanem2019} M. A. Ghanem, X. Liang, B. Lydon, L. Potocsnak,
  T. Wehr, M. Ghanem, S. Hoang, S. Cai, N. Boechler,
  Wrinkles riding waves in soft layered materials,
  Adv. Mat. Interface {\bf 6}, 1801609 (2019). 

\bibitem{Kodio2017} O. Kodio, I. M. Griffiths, D. Vella,
  Lubricated wrinkles: Imposed constraints affect the dynamics
  of wrinkle coarsening,
  Phys. Rev. Fluid {\bf 2}, 014202 (2017).

\bibitem{Chopin2017} J. Chopin, M. Dasgupta, A. Kudrolli,
  Dynamic wrinkling and strengthening of an elastic filament
  in a viscous fluid,
  Phys. Rev. Lett. {\bf 119}, 088001 (2017).

\bibitem{Box2020} F. Box, O. Kodio, D. O'Kiely, V. Cantelli, A. Goriely,
  D. Vella,
  Dynamic buckling of an elastic ring in a soap film,
  Phys. Rev. Lett. {\bf 124}, 198003 (2020).

\bibitem{Vandeparre2010} H. Vandeparre, S. Gabriele, F. Brau, C. Gay,
  K. K. Parker, P. Damman,
  Hierarchical wrinkling patterns,
  Soft Matter {\bf 6}, 5751--5756 (2010).

\bibitem{LLmechanics} L. D. Landau, E. M. Lifshitz, {\it Mechanics},
  2nd Ed., Pergamon Press (Oxford, 1960), sect. V.27. 

\bibitem{suppl} See Supplementary Material.

\bibitem{Groenewold2001} J. Groenewold, Wrinkling of plates coupled with
  soft elastic media, Physica A {\bf 298}, 32--45 (2001).

\bibitem{LLelasticityRayleigh} L. D. Landau, E. M. Lifshitz, {\it
  Theory of Elasticity}, 3rd Ed. (Butterworth-Heinemann, Oxford,
  1986), sect. III.24.

\bibitem{Lamb} H. Lamb, On the propagation of tremors over the surface
  of an elastic body, Phil. Trans. A {\bf 203}, 1--42 (1904).

\bibitem{LLelasticity} L. D. Landau, E. M. Lifshitz, {\it Theory of
  Elasticity}, 3rd Ed. (Butterworth-Heinemann, Oxford, 1986),
  sect. I.8.

\bibitem{Brau2011} F. Brau, H. Vandeparre, A. Sabbah, C. Poulard,
  A. Boudaoud, P. Damman, Multiple-length-scale elastic instability
  mimics parametric resonance of nonlinear oscillators,
  Nat. Phys. {\bf 7}, 56--60 (2011).

\bibitem{Brau2013} F. Brau, P. Damman, H. Diamant, T. A. Witten,
  Wrinkle to fold transition: influence of the substrate response,
  Soft Matter {\bf 9}, 8177--8186 (2013).

\bibitem{Dillard2018} D. A. Dillard, B. Mukherjee, P. Karnal,
  R. C. Batra, J. Frechette, A review of Winkler's foundation and its
  profound influence on adhesion and soft matter applications, Soft Matter
  {\bf 14}, 3669--3683 (2018).
  
\bibitem{SonnSegev2014} A. Sonn-Segev, A. Bernheim-Groswasser,
  H. Diamant, Y. Roichman, Viscoelastic response of a complex fluid at
  intermediate distances, Phys. Rev. Lett. {\bf 112}, 088301 (2014).

\bibitem{Grosberg2016} A. Y. Grosberg, J.-F. Joanny, W. Srinin, Y. Rabin,
  Scale-dependent viscosity in polymer fluids, J. Phys. Chem. B {\bf 120},
  6383--6390 (2016).
  
\bibitem{BarHaim2020} C. Bar-Haim, H. Diamant, Surface response of a
  polymer network: Semi-infinite network, Langmuir {\bf 36}, 247--255
  (2020).
    
\end{thebibliography}
\end{document}